\documentstyle[twocolumn,floats,ifthen,aps,prb]{revtex}

\begin{document}

\draft

\title{
Charge sensitivity of radio frequency single-electron transistor (RF-SET)
}

\author{Alexander N. Korotkov$^1$ and Mikko A. Paalanen$^2$}
\address{
$^1$Department of Physics, State University of New York, 
Stony Brook, New York 11794-3800 
\\ and 
$^2$Low Temperature Laboratory, Helsinki University of Technology, 
FIN-02015 HUT, Finland 
}

\date{\today}

\maketitle

\begin{abstract}
A theoretical analysis of the charge sensitivity of the RF-SET 
is presented. We use the ``orthodox'' approach and consider the case 
when the carrier frequency is much less than $I/e$ where $I$ is 
the typical current through RF-SET. The optimized noise-limited 
sensitivity is determined by the temperature $T$, and at low $T$
it is only 1.4 times less than the sensitivity of conventional 
single-electron transistor.     
\end{abstract} 


\narrowtext

\vspace{-0.2cm} 

        Single-electron devices \cite{Av-Likh} are gradually 
becoming useful in real applications.\cite{Kor-rev}
Despite the wide variety of studied circuits, the single-electron 
transistor (SET)\cite{Av-Likh,Likh-87,Fulton-D} remains the most 
important device in applied single-electronics 
(in this letter we will discuss the new version\cite{Schoelkopf}
of the SET setup). At present 
the best reported charge sensitivity of the SET at 10 Hz is\cite{Krupenin} 
$2.5\times 10^{-5} \, e/\sqrt{\mbox{Hz}}$ (the previous record
figure was\cite{Visscher} $7\times 10^{-5}\, e/\sqrt{\mbox{Hz}}$).
The low-frequency sensitivity of the SET is limited by 1/f noise, 
so it improves as the frequency increases. The best achieved 
so far\cite{Starmark} figure of 9$\times 10^{-6} \, e/\sqrt{\mbox{Hz}}$ 
was measured at 4.4 kHz. This is still an order of magnitude 
worse than the limit determined by the thermal/shot noise of the SET.
\cite{SQUID,Thesis,Hershfield,Galperin,Kor-sc,Kor-Lang} 

The difficulty of further frequency increase is due to the relatively 
large output resistance $R_d$ of the SET. For the typical figure 
$R_d\sim 10^5\, \Omega $  and wiring capacitance $C_L\sim 10^{-9}$ F 
the corresponding $R_d C_L$ time limits the bandwidth by a few kHz
(the use of filters can make it even lower).
The importance of potential high-frequency applications makes urgent a   
significant increase of the bandwidth. This can be done in several 
ways. 

The output resistance can be reduced in superconducting
(Bloch) SET based on supercurrent modulation \cite{Av-Likh,Zorin,Lukens} 
(the use of the quasiparticle tunneling threshold does not help much 
because  $R_d$ is limited by the quantum resistance 
even at the threshold \cite{Kor-sc,Av-Kor-PRL}). 
        The load capacitance $C_L$ can be decreased placing the next amplifier 
close to the SET. \cite{Pettersson,Visscher-2} However, while bandwidth 
up to 700 kHz was demonstrated\cite{Pettersson} using this idea,
the charge sensitivity was relatively poor because of  
extra heating and extra noise produced by the preamplifier. 
        Finally, a bandwidth over 100 MHz has recently been 
demonstrated\cite{Schoelkopf} in the so-called radio frequency (RF) SET 
in which the SET controlled the dissipation of the tank circuit which
in turn affected the reflection of the carrier wave with 
frequency $\omega/2\pi =1.7$ GHz. 
A sensitivity of $1.2\times 10^{-5} \, e/\sqrt{\mbox{Hz}}$
has been achieved\cite{Schoelkopf} at 1.1 MHz. The theoretical analysis 
of the ultimate sensitivity of the RF-SET is the subject of the present 
letter.

        In principle a wide bandwidth could be achieved 
simply by illuminating the SET with microwaves and measuring the
wave reflection. The gate voltage would change the SET
differential resistance $R_d$ and thus affect the reflection 
coefficient $\alpha =(Z-R_0)/(Z+R_0)$, where $Z^{-1}=i\omega C_s
+R_d^{-1}$,  $R_0\simeq 50\, \Omega$  is the cable wave resistance, and
$C_s$ is the stray capacitance. However, because of the large ratio 
$R_d/R_0 \sim 10^3$, the signal would be extremely small. To estimate
the signal power $P\simeq A^2R_0/2R_d^2[1+(\omega C_sR_0)^2]$, let us use
$R_d=10^5\, \Omega$ and the amplitude of the SET bias voltage oscillation 
$A=1$ mV ($A$ is limited by the Coulomb blockade threshold);
then $P\sim 10^{-15}$ W. This figure corresponds to the noise power 
of the amplifier with noise temperature of 10 K 
within 10$^7$ Hz bandwidth and clearly makes such an experiment 
quite difficult.

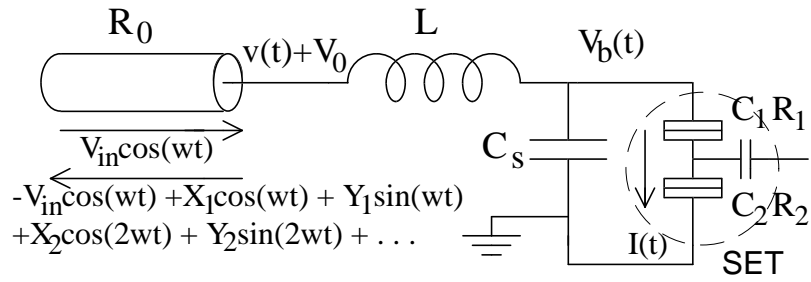
        \begin{figure}
\caption{The schematic of the RF-SET.}
\label{circuit}\end{figure}

        To increase the signal, the authors of Ref.\ \cite{Schoelkopf}
inserted the SET into the tank circuit (see Fig.\ \ref{circuit}).
Then at resonant frequency $\omega=(LC_s)^{-1/2}$ the circuit impedance
is small, $Z\simeq L/C_sR_d \ll R_0$ (we assume ${\cal Q}_{SET} \gg
{\cal Q} \gg 1$ where ${\cal Q}_{SET}= R_d/\sqrt{L/C_s}$ and ${\cal Q}=
\sqrt{L/C_s}/R_0$), so $\alpha \simeq -1+2L/C_sR_dR_0$. 
The signal power $P=[V_{in}(\alpha +1)]^2/2R_0$ ($V_{in}$ is
the amplitude of the incoming wave) can be expressed via the SET bias
amplitude $A\simeq 2{\cal Q} V_{in}$ as 
$P={\cal Q}^2A^2 R_0/2R_d^2$, indicating ${\cal Q}^2$ gain in comparison
with the nonresonant case.\cite{Schoelkopf} 

        The linear analysis above can be used only as an estimate
because of the considerable nonlinearity of the SET {\it I-V} curve.
For a more exact analysis let us write the differential equation
(see Fig.\ \ref{circuit}) for 
the voltage $v(t)$ at the end of the cable (the static component $V_0$
is subtracted):
\[ \ddot{v}LC_s +\dot{v}R_0C_s +v=2(1-\omega^2LC_s)V_{in}\cos \omega t -
R_0 I(t),  \]
where $V_{in} \cos \omega t$ is the incoming wave at the end of cable 
and $I(t)$ is the 
current through the SET while the SET bias voltage is 
$V_b(t)=V_0+v+(2V_{in}\omega \sin \omega t +\dot{v})L/R_0$. 
        The reflected wave can be written (at the end of cable) as 
$v(t)-V_{in}\cos \omega t = -V_{in}\cos\omega t + X_1 \cos\omega t
+Y_1 \sin\omega t +X_2\cos 2\omega t +Y_2\sin 2\omega t + \ldots$,
where the coefficients $X_k$ and $Y_k$ should be calculated self-consistently
(an obvious way is the iterative updating 
of $V_b(t)$ and $X_k,Y_k$). 
 While the analysis of the higher harmonics
is important for the possible versions of RF-SET in which the signal
is measured at the double (or triple) frequency, we will limit ourselves
by the reflected wave at the basic harmonic. 
For simplicity we assume exact resonance, $\omega =(LC_s)^{-1/2}$, then
        \begin{eqnarray}
X_1=2\sqrt{L/C_s}\, \langle I(t) \sin \omega t \rangle,
\nonumber \\
Y_1=2\sqrt{L/C_s}\, \langle I(t) \cos \omega t \rangle ,
        \end{eqnarray}
where $\langle \, \rangle$ denotes averaging over time. 
In the first approximation (if ${\cal Q}_{SET}\gg{\cal Q} \gg 1$) 
the SET bias voltage is 
$V_b(t)=V_0+A\sin \omega t$ where $A=2{\cal Q}V_{in}$.

The coefficients $X_1$ and $Y_1$ (we omit index 1 below) 
can be measured separately 
using homodyne detection and both can carry information about 
the low frequency signal applied to the SET gate (as usual,  
\cite{Av-Likh} we will describe it in terms of the background  
charge $Q_0$ induced into the SET island). 
        If the amplifier noise and other fluctuations are negligible,
then the sensitivity of the RF-SET is determined by the intrinsic
noise of the SET.
The minimal detectable charge $\delta Q$ can be expressed as 
        \begin{eqnarray}
\delta Q_X = \sqrt{S_{X}(f_s)\Delta f}/(dX/dQ_0), 
\nonumber \\
\delta Q_Y = \sqrt{S_{Y}(f_s)\Delta f}/(dY/dQ_0), 
        \label{dQ}\end{eqnarray} 
while the simultaneous measurement of $X$ and $Y$ can give 
$\delta Q =[(1-K^2)/(\delta Q_X^{-2}+\delta Q_Y^{-2} -2K/\delta Q_X
\delta Q_Y)]^{1/2}$, 
where $K=(\mbox{Re} S_{XY}/\sqrt{S_{X}S_{Y}})
\, \mbox{sign} [(dX/dQ_0)(dY/dQ_0)]$ is the 
correlation between two noises. Here $S_X(f_s)$ is the spectral density 
of $X(t)$ fluctuations at signal frequency $f_s$ (which should be within 
the tank circuit bandwidth, $2\pi f_s \lesssim \omega /{\cal Q}$), 
$S_{XY}$ is the mutual spectral density, and $\Delta f$ is the
measurement bandwidth (inverse ``accumulation'' time).

        In this letter we consider only the case of sufficiently
low carrier frequency $\omega \ll I/e$ (where $I$ is the typical current
through the SET), so that the quasistationary state is reached at any 
moment during the period of oscillations. In this case the spectral
density does not depend on $f_s$ (which is even lower than $\omega$) and
        \begin{equation} 
S_X=4(L/C_s)\, \langle S_I (t) \, \sin^2\omega t \rangle, 
        \end{equation} 
where $S_I (t)$ is the low frequency spectral density of the thermal/shot 
noise of the current 
through the SET, which has the time dependence because of oscillating
bias voltage $V_b$. There is no need to consider $Y$ output in this case 
because $Y =0$ (so $\delta Q_Y=\infty$) and 
the noise correlation is absent, $K=0$ (nonzero $Y$ and $K$ would appear 
at higher $\omega$ due to delay of tunneling events).

        We use the ``orthodox'' theory \cite{Av-Likh,Likh-87} for a 
normal SET 
consisting of two tunnel junctions with capacitances $C_1$ and $C_2$ 
and resistances $R_1$ and $R_2$ (see Fig.\ \ref{circuit}) assuming 
$R_j \gg R_Q=\pi \hbar/2e^2$ (as usual, the gate 
capacitance is distributed between $C_1$ and $C_2$ in a proper way).
The effects of finite photon energy $\hbar \omega$ are neglected.
We also neglect the possible rf modulation of the SET gate voltage.
The low frequency thermal/shot noise of the SET current is calculated
in the standard way. \cite{SQUID,Thesis}
        
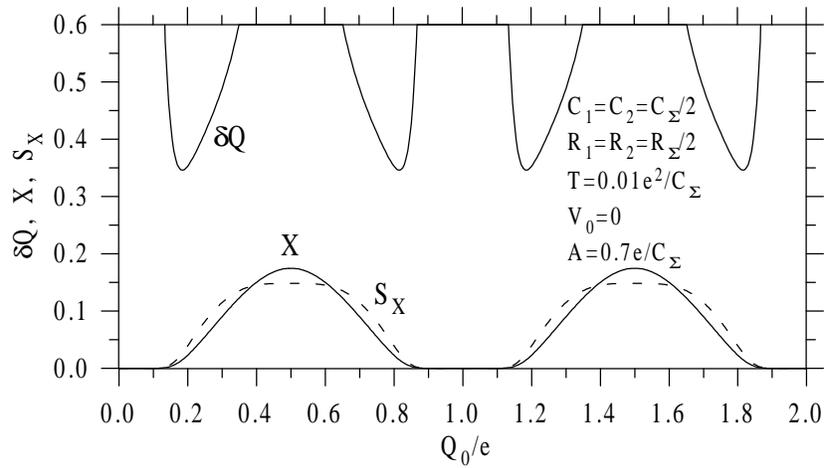
        \begin{figure}
\caption{The reflected wave amplitude $X_1$ (in units 
        $2(L/C_s)^{1/2} e/R_\Sigma C_\Sigma$), its noise $S_{X1}$ 
        (in units $4 L/C_s e^2/R_\Sigma C_\Sigma$), 
        and minimal detectable charge $\delta Q$ 
        (in units $e(R_\Sigma C_\Sigma\Delta f)^{1/2}$) as functions of
	the background charge
        $Q_0$ for the symmetric SET at $T=0.01 e^2/C_\Sigma$ for 
        zero dc bias voltage $V_0$ and its rf amplitude 
        $A=0.7 e/C_\Sigma$.} 
\label{fig2}\end{figure}

        Figure \ref{fig2} shows the dependence of $X$, $S_{X}$,
and $\delta Q=\delta Q_X$ on the background charge $Q_0$ for a 
symmetric SET ($C_1=C_2$, $R_1=R_2$) at $T=0.01e^2/C_\Sigma$ 
($C_\Sigma =C_1+C_2$), $V_0=0$, and $A=0.7e/C_\Sigma$. 
One can see that the minimum of $\delta Q$ is achieved near the edge
of $Q_0$ range corresponding to nonzero $X$, so that the amplitude 
$A$ is only a little larger 
than the Coulomb blockade threshold $V_t$. For $V_b$ close to $V_t$ 
the noise of the
current through the SET obeys Schottky formula, $S_I=2eI$, with 
a good accuracy at low temperatures,\cite{SQUID,Thesis} while
the current $I$ can be approximated as $I=W/eR_j[1-\exp(-W/T)]$ 
where $W=e(V_b-V_t)(C_1C_2/C_jC_\Sigma)=(-1)^j e(Q_0-Q_{0,t})/C_\Sigma$ 
($j$th junction determines the threshold) and 
$|dI/dQ_0|=(dI/dV_b)C_j/C_1C_2$. (As a consequence of the Schottky
formula, the dashed curve in Fig.\ \ref{fig2} is approximately 
twice as high as the $X$-curve at small $X$.) 

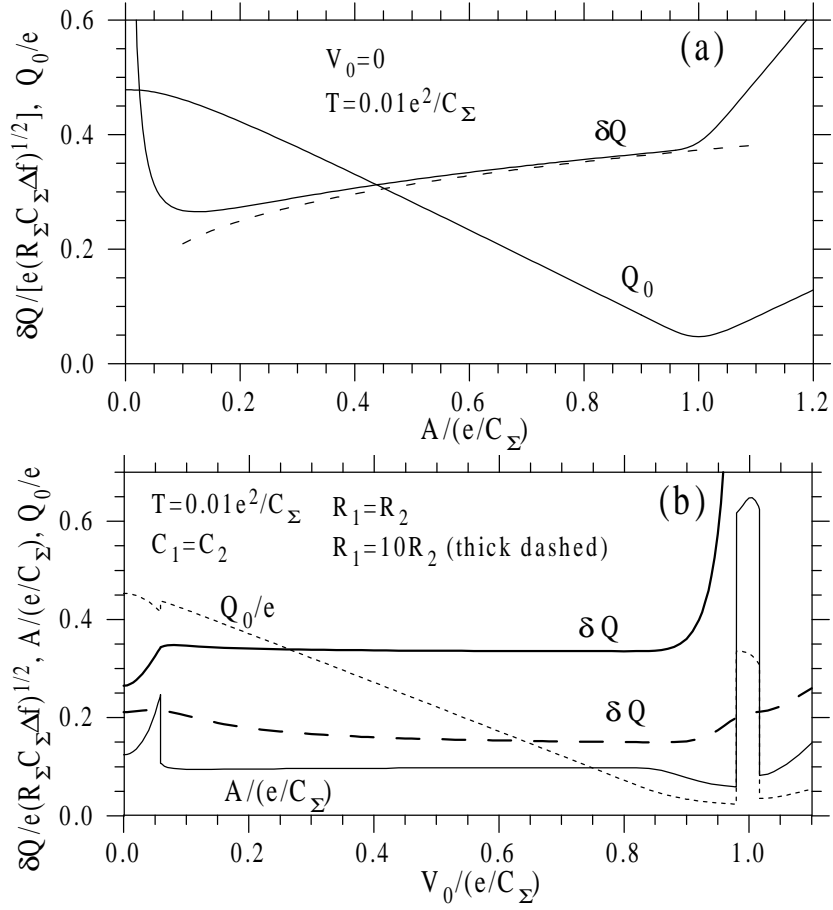
        \begin{figure}
\caption{(a) The sensitivity $\delta Q$ optimized over $Q_0$ 
        and corresponding $Q_0$ as functions of rf amplitude $A$ for
        the symmetric SET. Dashed line shows the analytical result 
        (see text). 
        (b) Dependence of $\delta Q$ minimized over $A$ and $Q_0$ 
        and of the optimal operation point ($A,Q_0$) 
	on the dc bias voltage 
	$V_0$. Dashed line is for the asymmetric SET ($R_2/R_1=10$).} 
\label{fig3}\end{figure}

Using these equations and optimizing $Q_0$, one can find the minimum 
$\delta Q \simeq 1.2 \, e \, (R_\Sigma C_\Sigma \Delta f)^{1/2} 
(TC_\Sigma/e^2)^{1/2} \times (eA/T)^{1/4}$ for the symmetric SET 
at $T\ll eA < e^2/C_\Sigma$ ($R_\Sigma =R_1+R_2$). 
This dependence as a function of rf amplitude $A$ is shown in 
Fig.\ \ref{fig3}a by the dashed line while the numerical result 
is shown by the solid line. The sensitivity gets worse ($\delta Q$
increases) at $A>e/C_\Sigma$ 
because of $X$ and $S_X$ increase. The sensitivity also worsens 
rapidly when $A$ is too small and becomes comparable to $T/e$, 
because of the contribution from the Nyquist noise of the SET 
at $V_b$ close to zero. Before optimizing the amplitude $A$, let us 
notice that the results shown in Fig.\ \ref{fig3}a correspond 
to relatively small $X$ that can be difficult to measure  
experimentally [in the approximation above  
$X\simeq 2\, (L/C_s)^{1/2}\times 15\, (T/eR_\Sigma)(T/eA)^{1/2}$]. 
However, as seen from Fig.\ \ref{fig2}, $X$ can be significantly
increased for the price of a few ten per cent increase of $\delta Q$.

        Figure \ref{fig3}b shows $\delta Q$ minimized over both $A$ and 
$Q_0$ and the corresponding optimum values of $A$ and $Q_0$ as functions 
of the dc bias voltage $V_0$.
One can see that for a symmetric SET the best sensitivity is 
achieved at $V_0=0$ and there is a long plateau of $\delta Q$ which ends
when $V_0$ approaches $e/C_\Sigma$ leading to significant worsening
of the sensitivity. 
For the asymmetric SET (dashed line) the best sensitivity can be
achieved in the plateau range. 
At the plateau $\delta Q$ can be calculated analytically using 
the approximations above,  $\delta Q \simeq 3.34\, e\,  
(2R_{min} C_\Sigma\Delta f)^{1/2} (TC_\Sigma /e^2)^{1/2}$ where 
$R_{min}=\mbox{min}(R_1,R_2)$. This expression can be compared 
with the optimized low-temperature sensitivity of the conventional
SET which is given\cite{SQUID,Thesis} by the same formula with
the numerical factor 1.90 instead of 3.34. 
For the symmetric RF-SET the optimized low-temperature sensitivity
(at $V_0=0$) is
        \begin{equation}
\delta Q \simeq 2.65\, e\, (R_\Sigma C_\Sigma\Delta f)^{1/2} 
(TC_\Sigma /e^2)^{1/2},  
        \label{dq-opt}\end{equation}
only 1.4 times worse than for the conventional SET. 

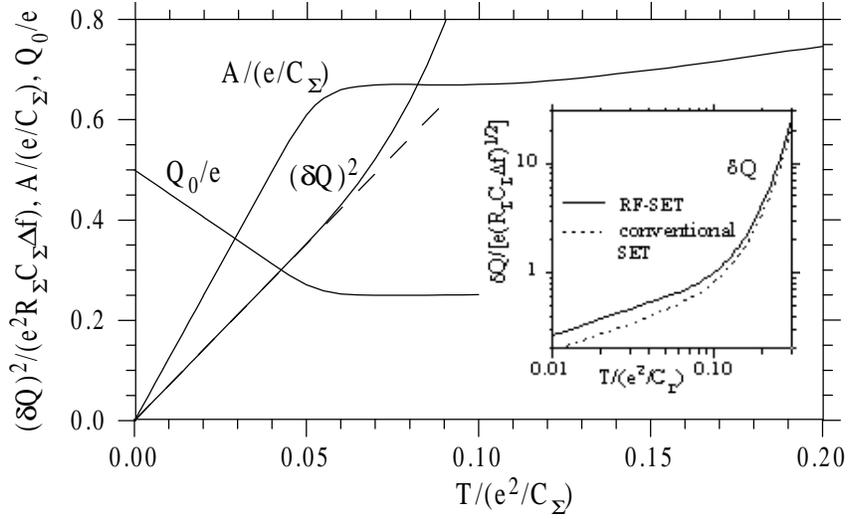
        \begin{figure}
\caption{The optimized $\delta Q$ (squared) and corresponding $A$ and $Q_0$
        as functions of the temperature $T$ for the symmetric SET at 
        $V_0=0$. Dashed line represents Eq.\ (\protect\ref{dq-opt}). 
        Inset shows $\delta Q$ on the larger scale. For comparison, the
        result for conventional SET is shown by the dotted line.} 
\label{fig4}\end{figure}

        Figure \ref{fig4} shows numerically minimized $\delta Q$
for the symmetric SET and corresponding optimal 
$A$ and $Q_0$ (while $V_0=0$) as functions of temperature. 
The result of Eq.\ (\ref{dq-opt}) 
is shown by the dashed line. The sensitivity scales as $T^{1/2}$ at low 
temperatures while it significantly worsens at $T>0.1e^2/C_\Sigma$, 
similar to the result for the conventional SET (dotted line). 
	The ``orthodox'' sensitivity improves with the decrease of tunnel  
resistances while the optimum value (which should be comparable
to $R_Q$) could be calculated if cotunneling \cite{Av-Likh} 
was taken into account. 

	To make a comparison with experiment, \cite{Schoelkopf}
let us take $C_\Sigma =0.45$ fF, $R_\Sigma=97$ k$\Omega$, and
$T=100$ mK, then after optimization $\delta Q\simeq 2.7\times 10^{-6} \, 
e/\sqrt{\mbox{Hz}}$ in the normal case (necessity of relatively large 
$X$ would lead to a factor about 1.5). So, there is still an order 
of magnitude for possible experimental improvement.
Comparison for the 
superconducting case is not straightforward because the sensitivity 
depends on the junction quality.\cite{Kor-sc} 
        
        In conclusion, we have shown that the price for the wide 
bandwidth of the RF-SET is only a little decrease of the 
noise-limited sensitivity in comparison with conventional SET. 
        The authors thank K.\ K.\ Likharev and H. Sepp\"a for
valuable discussions. 
        The work was supported in part by US AFOSR, Russian Fund for 
Basic Research, and Finnish Academy of Sciences and Letters. 


\end{document}